\documentclass[%
 reprint,
 prc,
 amsmath,amssymb,
 aps,
 lengthcheck,
floatfix,
]{revtex4}

\usepackage{graphicx}
\usepackage{dcolumn}
\usepackage{bm}
\usepackage{hyperref}
\usepackage[utf8]{inputenc}
\usepackage{multirow}
\usepackage{soul}
\usepackage{float}
\usepackage{graphicx}
\usepackage{times}
\usepackage[normalem]{ulem}
\usepackage{color}
\usepackage{cellspace}
\usepackage{lipsum}

\newcommand{\be}{\begin{equation}}
\newcommand{\ee}{\end{equation}}
\newcommand{\bea}{\begin{eqnarray}}
\newcommand{\eea}{\end{eqnarray}}

\newcommand{\comment}[1]{}
\renewcommand\sout{\bgroup \color{red} \ULdepth=-.5ex \ULset}
\def\simge{\mathrel{\rlap{\raise 0.511ex
     \hbox{$>$}}{\lower 0.511ex \hbox{$\sim$}}}}
\def\simle{\mathrel{\rlap{\raise 0.511ex
      \hbox{$<$}}{\lower 0.511ex \hbox{$\sim$}}}}

\usepackage[dvipsnames]{xcolor}

\begin{document}


\title{Effect of the $\sigma$-cut potential on the properties of neutron stars with or without a hyperonic core}

%

\author{N. K. Patra$^1$}
\author{B. K. Sharma$^2$}
\email{bk\_sharma@cb.amrita.edu }
\author{A. Reghunath$^{2,3}$}
\author{A. K. H. Das$^2$}
\author{T. K. Jha$^1$}

\affiliation{$^1$Department of Physics, BITS-Pilani, K. K. Birla Goa Campus, Goa 403726, India.}
\affiliation{$^2$Department of Sciences, Amrita School of Physical Sciences, 
Amrita Vishwa Vidyapeetham, Coimbatore 641112, India}
\affiliation{$^3$ Institut fuer Physik, Experimentelle Elementarteilchenphysik II (EE2),
Humboldt-Universitaet zu Berlin, 12489 Berlin, Germany}

\date{\today}

\begin{abstract} 
Motivated by the recent observation of high-mass pulsars ($M \simeq 2 M_{\odot}$), we employ the $\sigma$-cut potential on the equation of state (EOS) of high-density matter and the properties of neutron stars within the relativistic mean-field (RMF) model using TM1$^{*}$ parameter set. The $\sigma$-cut potential is known to reduce the contributions of the $\sigma$ field, resulting in a stiffer EOS at high densities and hence leading to larger neutron star masses without affecting the properties of nuclear matter at normal saturation density. We also analyzed the effect of the same on pure neutron matter and also on the neutron star matter with and without hyperonic core and compared it with the available theoretical, experimental, and observational data. The corresponding tidal deformability ($\Lambda_{1.4}$) is also calculated. With the choice of meson-hyperon coupling fixed to hypernuclear potentials, we obtain $\approx 10~\%$ increase in mass by employing the $\sigma$-cut potential for $f_{s} = 0.6$. Our results are in good agreement with various experimental constraints and observational data, particularly with the GW170817 data.
\end{abstract}


\maketitle
\section{Introduction}
Neutron stars (NSs) are one of the densest objects that exist in the universe
\cite{Baym:1975mf, Baym:1978jf, Heiselberg:2000dn}. All four fundamental forces play
an important role in determining the global properties of these stars.
The study of their global properties provides a crucial link between nuclear,
particle, and astrophysics \cite{Malik2018, Patra2020, Patra:2022yqc}. The neutron star matter is likely to be composed of ions
and electrons in the outer crust region, i.e., near the surface,
and neutron-rich nuclei and some free neutrons appear in the inner crust region. In the core densities, it may have free neutrons along with fewer protons and electrons.
As we go deeper to the core of the star, i.e., with an increase of density towards the center of the star, exotic components of NS matter, viz., hyperons, heavier
non-strange baryons, boson condensates, and even deconfined
quarks may appear\cite{Gledenning1996}.
The global properties of the neutron star
are the imprints of a particular equation of state (EOS). Energetically, the hyperons start appearing via weak interaction around 2-3 times nuclear saturation density $\rho_{0}$.
The presence of hyperons makes EOS softer and consequently predicts a mass lower than when they are not included and has an effect on the radius too. In addition, Softer EOS at higher density may also be achieved by considering cross couplings among the $\sigma$, $\omega$, and $\rho$
mesons within the Relativistic Mean Field approach \cite{DelEstal:2001qr,Jha2017}. 

A comprehensive idea regarding the $\sigma$-cut scheme was convened by studying its effects on neutron star properties such as maximum mass, compactness, and tidal deformability.
Recently, two different groups of Neutron Star Interior Composition Explorer (NICER) X-ray telescopes provided simultaneously  neutron star's mass and radius  
for PSR J0030+0451 with
$R(1.44^{+0.15}_{-0.14})=13.02^{+1.24}_{-1.06}$km \cite{Miller:2019cac} and
$R(1.34^{+0.15}_{-0.16})=12.71^{+1.14}_{-1.19}$km \cite{Riley:2019yda}
and for J0740+6620 with
$R(2.08 \pm 0.07)=13.7^{+2.6}_{-1.5}$km \cite{Miller:2021qha} and
$R(2.072^{+0.067}_{-0.066})=12.39^{+1.30}_{-0.98}$km \cite{Riley:2021pdl}.
Due to the model dependence of the experimental analyses, differences between these estimates came out. By combining the NICER results with the limits on the NS maximum mass
as well as the tidal deformability from GW170817 \cite{Abbott2020}
leads to constraints on the $\beta$-equilibrated EOS
for densities in the range $1.5 \rho_0 \lesssim \rho_B \lesssim 3\rho_0$ \cite{Miller:2021qha,Raaijmakers:2021uju},
which can limit NS mass and radii
\cite{Miller:2021qha,Riley:2019yda,Miller:2019cac,Riley:2021pdl,Pang:2021jta,Raaijmakers:2021uju}.

Observation of neutron star's mass around 2$M_{\odot}$ \cite{Demorest:2010bx,Antoniadis:2013pzd} and the corresponding radius estimates around 11.9 $\pm$ 1.4 km \cite{Abbott2018,Abbott2019} respectively disfavors the possibility of softer EOS. On the contrary, results obtained from transport models predict a soft EOS \cite{Fuchs:2000kp, Xiao:2008vm} at high densities. Such apparent contradictions in the EOS from experimental and observational data motivate one to probe the different aspects of an EOS and emphasize the importance to constrain them \cite{Chatterjee:2015pua}.

A $\sigma$-cut scheme \cite{Maslov:2015lma} was developed to make EOS stiffer
at high densities without compromising the properties of nuclear matter around
saturation density $\rho_{0}$. In the $\sigma$-cut scheme, a $\sigma$-potential term is
included in the Lagrangian of the RMF model to reduce the contribution of the scalar field
at high density thereby making EOS stiffer resulting in larger masses for neutron stars. { In one of recent works \cite{Dutra:2015hxa}, a plethora of RMF models were analyzed to test the range of resulting neutron star masses, where 14 of them could result in masses within the range (1.93 - 2.05)$M_{\odot}$, of which only two of the models could satisfy the mass constraint when hyperons were included. To increase the masses further, in accordance with recent observations, one of the ways out is to include the $\sigma$-cut scheme and analyze the effect on the EOS as well as the star masses.} Recently, the $\sigma$-cut scheme is implemented to study the properties of finite nuclei and
hyperon-rich matter using TM1 parameter set \cite{Zhang:2018lpl} and also to
kaon condensate in neutron stars \cite{Ma:2022fmu} using FSUGold parameter set.

In the present work, we implement the $\sigma$-cut scheme using TM1$^{*}$ \cite{DelEstal:2001yz} interaction.
The improved TM1$^{*}$ parameter has extra cross couplings compared to TM1. These extra couplings have no effect on the
properties of finite nuclei and nuclear matter around saturation density but these cross couplings seem to be instrumental in making EOS softer at high densities. We choose TM1$^{*}$ for the present analysis and use the same strength of $\sigma$-cut potential as given in ref. \cite{Zhang:2018lpl}.
       
The paper is organized as follows: In Sec. II, we briefly describe the RMF
model with $\sigma$-cut potential and stellar equations for a neutron star.
In Sec. III, we present and discuss the role of $\sigma$-cut potential on the properties
of nuclear matter and neutron stars with hyperons. The summary and conclusions of the present
work are given in Sec. IV.                                                                

\section{Formalism}
We apply the well-known RMF model to describe the EOS
at higher densities relevant to the neutron stars. The details of the model
and derivation of EOS with nucleon and with nucleon and hyperon both can be found
in the ref. \cite{Sharma:2006nw}. We include the $\sigma$-cut potential $U_{cut}(\sigma)$ in
RMF Lagrangian as in the refs.\cite{Maslov:2015lma,Zhang:2018lpl}.
The Lagrangian density with $U_{cut}(\sigma)$ is given by,

\begin{eqnarray}
{\cal L} & = & \sum_{B}\overline{\Psi}_{B}\left ( i\gamma^\mu
D_{\mu} - M_{B} + g_{\sigma B}{\sigma}\right)
{\Psi}_{B}
+\frac{1}{2}{\partial_{\mu}}{\sigma}{\partial^{\mu}}{\sigma}
\nonumber \\
&&-m_{\sigma}^2{\sigma^2}\left(\frac{1}{2}
+\frac{\kappa_3}{3 !}
\frac{g_{\sigma B}\sigma}{M_{N}}+\frac{\kappa_4}{4 !}
\frac{g^2_{\sigma B}\sigma^2}{M_{N}^2}\right)
-\frac{1}{4} {\Omega_{\mu\nu}}{\Omega^{\mu\nu}}
\nonumber \\
&&+\frac{1}{2}\left (1 +
{\eta_1}\frac{g_{\sigma B}\sigma}{M_{N}}
+\frac{\eta_2}{2}\frac{g^2_{\sigma B}\sigma^2}{M_{N}^2}\right)
m_{\omega}^2{\omega_{\mu}}
{\omega^{\mu}}
- \frac{1}{4}{R^a_{\mu\nu}}{R^{a\mu\nu}}
\nonumber \\
&&+\frac{1}{2}\left(1 + \eta_{\rho}\frac{g_{\sigma B}\sigma}{M_{N}}\right)
m_{\rho}^2{\rho^a_{\mu}}{\rho^{a\mu}}+\frac{1}{4 !}{\zeta_{0}}
g^2_{\omega B}\left({\omega_{\mu}}{\omega^{\mu}}\right)^2
\nonumber \\
&&-U_{cut}(\sigma)
\label{lag}
\end{eqnarray} 

The $U_{cut}(\sigma)$ has logarithmic form as \cite{Maslov:2015lma} which only influence the
$\sigma$-field at high density and is given by, 
 
\begin{eqnarray}
U_{cut}(\sigma) = \alpha \ln [ 1 + \exp\{\beta(g_{\sigma N}\sigma/M_{N}-f_{s})\}]
\end{eqnarray}
where $\alpha$ = $m_{\pi}^{4}$ and $\beta$ = 120 \cite{Maslov:2015lma}.
The factor $f_{s}$ is a free parameter and we take $f_{s}$ = 0.6 \cite{Zhang:2018lpl} for
our calculation. The field equations for $\sigma$, $\omega$, and $\rho$-mesons obtained from
eq.(1) are given by,

\begin{eqnarray}
m_{\sigma}^{2}\left({\sigma}_{0} + \frac{g_{\sigma N}{\kappa_{3}}}{2M_{N}}\sigma_{0}^{2}
+ \frac{g_{\sigma N}^{2}{\kappa_{4}}}{6M_{N}^{2}}\sigma_{0}^{3}\right) + U_{cut}^{'}(\sigma)\nonumber\\
-\frac{1}{2}m_{\omega}^{2}\left(\eta_{1}\frac{g_{\sigma N}}{M_{N}} + \eta_{2}\frac{g_{\sigma N}^{2}}
{M_{N}^{2}}\sigma_{0}\right)\omega_{0}^{2}\nonumber\\
-\frac{1}{2}m_{\rho}^{2}\eta_{\rho}\frac{g_{\sigma N}}{M_{N}}
\rho_{0}^{2} 
= \sum_{B} g_{\sigma B}M_{B}^{*2}\rho_{SB}
\end{eqnarray}

\begin{eqnarray}
m_{\omega}^{2}\left(1 + \frac{\eta_{1}g_{\sigma N}}{M_{N}}\sigma_{0}
+ \frac{\eta_{2}g_{\sigma N}^{2}}{2M_{N}^{2}}
\sigma_{0}^{2}\right)\omega_{0}\nonumber\\ 
+ \frac{1}{6}\zeta_{0}g_{\omega N}^{2}\omega_{0}^{3} = \sum_{B} g_{\omega B}\rho_{B}
\end{eqnarray}

\begin{eqnarray}
m_{\rho}^{2}\left(1 + \frac{g_{\sigma N}{\eta_{\rho}}}{M_{N}}\sigma_{0}\right)\rho_{03}
= \sum_{B} g_{\rho B}I_{3B}\rho_{B}
\end{eqnarray}
where $I_{3B}$ is the third component of nucleon isospin operator
and the derivative of $U_{cut}(\sigma)$ is given by,
\begin{eqnarray}
U_{cut}^{'}(\sigma) = \frac{\alpha\beta g_{\sigma N}}{M_{N}}
 \frac{1}{[ 1 + \exp\{-\beta(g_{\sigma N}\sigma/M_{N}-f_{s})\}]}
\end{eqnarray}
The EOS for hyperon-rich matter satisfies the conservation of the total baryon number and charge
neutrality condition which is given by,
\begin{eqnarray}
\sum_{B} Q_{B}\rho_{B} + \sum_{L} Q_{L}\rho_{L} = 0
\end{eqnarray}

where $\rho_{B}$ and $\rho_{L}$ are the baryon and the lepton $(e, \mu)$ number densities with $Q_{B}$ and $Q_{L}$ as their respective electric charge.

The energy density $\cal E$ and pressure $P$ for charge-neutral $\beta$-equilibrated neutron star matter
with a lowest-lying octet of baryons is given by,
\begin{eqnarray}
{\cal E} &=& \sum_{B} \frac{2}{(2\pi)^{3}}\int_{0}^{k_{B}}d^{3}k{E_{B}^{*}(k)}
+ \frac{1}{8}\zeta_{0}g_{\omega N}^{2}
\omega_{0}^{4} + U_{cut}(\sigma)\nonumber\\ 
&+& \frac{1}{2}\left(1 + \frac{\eta_{1}g_{\sigma N}}{M_{N}}\sigma_{0}
+ \frac{\eta_{2}g_{\sigma N}^{2}}{2M_{N}^{2}}\sigma_{0}^{2}\right)m_{\omega}^{2}{\omega_{0}^{2}}\nonumber\\
&+& m_{\sigma}^{2}{\sigma_{0}}^{2}\left(\frac{1}{2} + \frac{\kappa_{3}g_{\sigma N}\sigma_{0}}{3!M_{N}}
+ \frac{\kappa_{4}g_{\sigma N}^{2}\sigma_{0}^{2}}{4!M_{N}^{2}}\right)\nonumber\\
&+& \frac{1}{2}\left(1 + \eta_{\rho}\frac{g_{\sigma N}\sigma_{0}}{M_{N}}\right)m_{\rho}^{2}{\rho_{0}}^{2}
+ \sum_{L} {\cal E}_{L}
\end{eqnarray}

\begin{eqnarray}
P &=& \sum_{B} \frac{2}{3(2\pi)^{3}}\int_{0}^{k_{B}}d^{3}k\frac{k^{2}}{E_{B}^{*}(k)}
+ \frac{1}{4!}\zeta_{0}g_{\omega N}^{2}
\omega_{0}^{4} - U_{cut}(\sigma)\nonumber\\ 
&+& \frac{1}{2}\left(1 + \frac{\eta_{1}g_{\sigma N}}{M_{N}}\sigma_{0}
+ \frac{\eta_{2}g_{\sigma N}^{2}}{2M_{N}^{2}}\sigma_{0}^{2}\right)m_{\omega}^{2}{\omega_{0}^{2}}\nonumber\\
&-& m_{\sigma}^{2}{\sigma_{0}}^{2}\left(\frac{1}{2} + \frac{\kappa_{3}g_{\sigma N}\sigma_{0}}{3!M_{N}}
+ \frac{\kappa_{4}g_{\sigma N}^{2}\sigma_{0}^{2}}{4!M_{N}^{2}}\right)\nonumber\\
&+& \frac{1}{2}\left(1 + \eta_{\rho}\frac{g_{\sigma N}\sigma_{0}}{M_{N}}\right)m_{\rho}^{2}{\rho_{0}}^{2} + \sum_{L} P_{L}
\label{pden}
\end{eqnarray}
Here the subscript 'B', 'N', and 'L' represents the low-lying octet of baryons, nucleons, and leptons
respectively. The ${\cal E}_{L}$ and $P_{L}$ are the energy density and pressure of the leptons.

The mass-radius relation for a neutron star is obtained by Tolman, Oppenheimer, and Volkoff (TOV) equation \cite{Oppenheimer:1939ne, Tolman:1939jz}
which is given by,
\begin{equation}
\frac{dP}{dr}=-\frac{G}{r}\frac{\left[\varepsilon+P\right ]
\left[M+4\pi r^3 P\right ]}{(r-2 GM)},
\label{tov1}
\end{equation}
\begin{equation}
\frac{dM}{dr}= 4\pi r^2 \varepsilon,
\label{tov2}
\end{equation}
\noindent

Here we adopt the natural units, i.e., c = 1, and the terms G, $P(r)$ and $M(r)$ are the universal
gravitational constant, the pressure of neutron star, and the enclosed gravitational mass
inside a sphere of radius $(r)$ respectively.
The Eqs. (\ref{tov1}) and (\ref{tov2}) are solved to obtain the structural properties of a
static neutron star with neutral hyperonic matter \cite{Lattimer:2004pg,Krastev:2006ii}.

\section{Results and Discussion}
We choose TM1$^{*}$ parameter set \cite{DelEstal:2001yz} for our present analysis. As already mentioned, the motivation to choose TM1$^{*}$ is to examine how the $\sigma$-cut potential
influence the EOS with cross-couplings of $\sigma$-field and consequently the structural properties
of a neutron star. In $\sigma$-cut scheme with TM1$^{*}$ we took $\alpha$ = $m_{\pi}^{4}$ and $\beta$ = 120 as given in \cite{Maslov:2015lma} except the
value of factor $f_{s}$. In ref. \cite{Maslov:2015lma} the value of $f_{s}$ was taken as 0.36, 0.44 and 0.52
unlike \cite{Zhang:2018lpl}. In one of the earlier works, for finite nuclei results including $\sigma$-cut potential are found to be identical with the original TM1 parameter
set for the value of $f_{s}$ larger than 0.55 \cite{Zhang:2018lpl}. In the finite nuclear domain, there is almost no difference between TM1 and TM1$^{*}$ \cite{DelEstal:2001yz} and we choose $f_{s}$ = 0.6 for our present analysis.  

\begin{figure}
\includegraphics[width=8cm,height=8cm,angle=0]{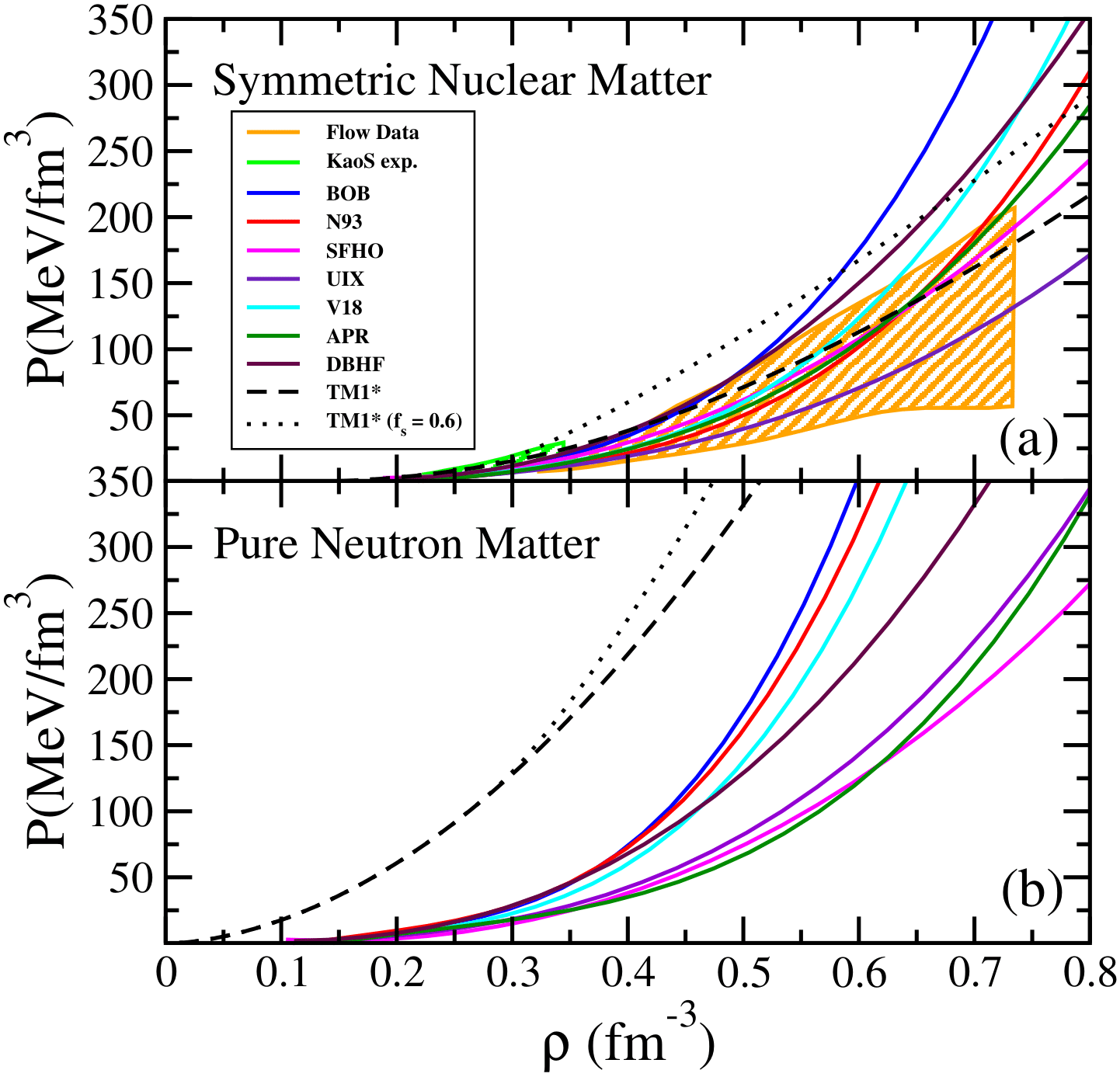}
\caption{Pressure as a function of number density $\rho$ for (a) symmetric nuclear matter
and (b) pure neutron matter with original TM1$^{*}$ and the TM1$^{*}$ with $\sigma$-cut potential ($f_{s}$ = 0.6) along with other microscopic EOSs. The orange polygon and green polygon in Fig.1(a) 
describe the transverse flow data \cite{FOPI:1995swp} and KaoS experimental data \cite{Miskowiec:1994vj} respectively.} 
\label{fig1}
\end{figure}

In the present analysis, we used several microscopic Brueckner–Hartree–Fock (BHF) EOS based on different 
nucleon-nucleon potential namely Bonn B (BOB) \cite{Machleidt:1987hj,Machleidt:1989tm}, the Nijmegen 93 (N93) \cite{Nagels:1977ze,Stoks:1994wp},
Argonne $V_{18}$ (V18) \cite{Wiringa:1994wb} as well as phenomenological Urbana model (UIX), Dirac-BHF method (DBHF) 
\cite{Gross-Boelting:1998xsk}, APR EOS \cite{Akmal:1998cf} and well-known phenomenological EOS SFHO \cite{Steiner:2012rk} for symmetric
nuclear matter, pure neutron matter, and symmetry energy for comparison along with available experimental data.   

In Fig.1, we plot and compare all the microscopic EOSs along with the original TM1$^{*}$ and TM1$^{*}$ with $\sigma$-cut 
potential ($f_{s}$ = 0.6) for symmetric nuclear matter and pure neutron matter.
In Fig.1(a), the orange and green shaded regions represent the flow data by the FOPI collaboration \cite{FOPI:1995swp}
and data from KaoS collaboration \cite{Miskowiec:1994vj} for symmetric nuclear matter, respectively.
From Fig.1(a), we see that most of the microscopic EOSs are compatible with the experimental data \cite{FOPI:1995swp, Miskowiec:1994vj}
except for BOB, V18, and DBHF EOS, which are stiff at high densities. It is to be noted that the original TM1$^{*}$ results are 
consistent with experimental data at all the values of number density $\rho$. On the other hand, the TM1$^{*}$ EOS with 
$\sigma$-cut potential gives the stiffest EOS for symmetric nuclear matter compared to other EOSs. 
The EOS with $\sigma$-cut potential becomes stiffer at $\rho$ = $0.28$ $fm^{-3}$ compared to original 
TM1$^{*}$ EOS for symmetric nuclear matter. 
For completeness, we plot all considered EOSs for pure neutron matter in fig.1(b). 
We find that the original TM1$^{*}$ and EOS with $\sigma$-cut potential is stiffer compared to other
microscopic EOSs for pure neutron matter. The TM1$^{*}$ EOS with
$\sigma$-cut potential becomes stiffer at $\rho$ = $0.32$ $fm^{-3}$ compare to original TM1$^{*}$ EOS for pure neutron 
matter. One can conclude from fig.1 that the effect of $\sigma$-cut potential is more in symmetric nuclear matter compared
to pure neutron matter. The reason for this discrepancy is that the $\sigma$-cut potential reduces the magnitude of $\sigma$-field contribution in symmetric nuclear matter and $\omega$-field remains unaffected 
by $\sigma$-cut potential resulting in stiffer EOS stiffer at high densities. On the other hand,
the contribution of $\sigma$-field is smaller in pure neutron matter is smaller compared to symmetric nuclear matter. 
As $\sigma$-cut potential is isospin independent and equally effect $\sigma$-field in symmetric nuclear and pure neutron
matter. The contribution of $\sigma$-field is smaller in pure neutron matter compared to the symmetric nuclear matter at
certain densities without $\sigma$-cut potential. So, the effect of $\sigma$-cut potential is less prominent in neutron
matter as compared to the symmetric nuclear matter. 

\begin{figure}
\includegraphics[width=9.0cm,height=8cm,angle=0]{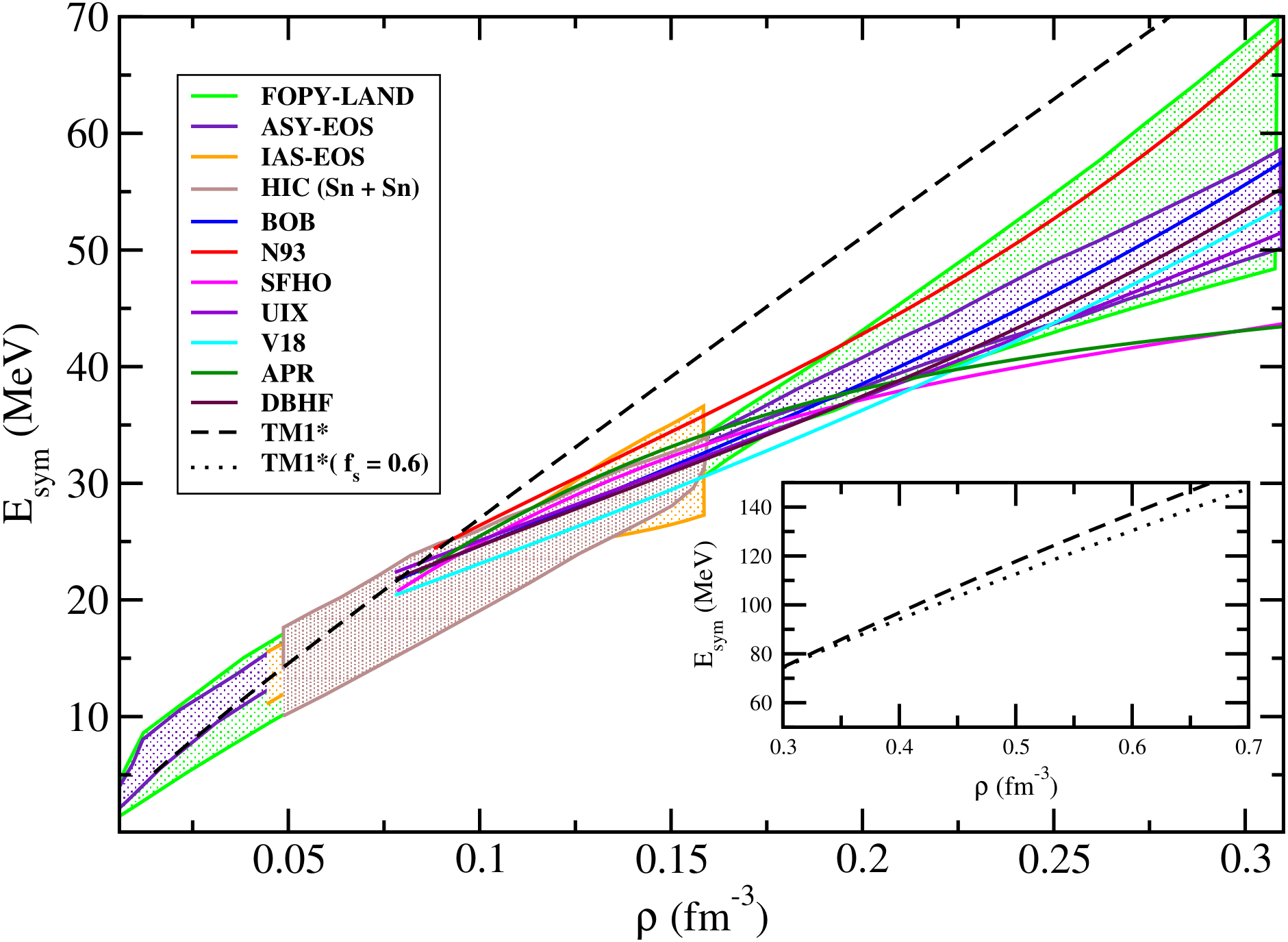}
\caption{Symmetry energy $E_{sym}$ as a function of number density $\rho$ for all microscopic EOS as well as with original TM1$^{*}$ and the TM1$^{*}$ with $\sigma$-cut potential along with experimental data. The inset plot shows $E_{sym}$ for the original TM1$^{*}$ and the TM1$^{*}$
with $\sigma$-cut potential at high densities.} 
\label{fig2}
\end{figure}

The symmetry energy $E_{sym}$ and its density dependence play an important role in finite nuclei and
neutron star properties. A lot of experimental efforts have been put into constraining high-density
behavior of $E_{sym}$ using various probes in heavy ion collisions (HICs) at relativistic energies.
In Fig.2, we display some constraints obtained for the density dependence of $E_{sym}$ from FOPY-LAND (green band)
\cite{Russotto:2011hq}, ASY-EOS (indigo band) \cite{Russotto:2016ucm}, IAS-EOS (orange band) \cite{Tsang:2008fd}, and HIC (Sn + Sn) (brown band)
\cite{Danielewicz:2013upa} respectively. We also plot all the microscopic EOSs along with the original TM1$^{*}$ and the TM1$^{*}$
$\sigma$-cut potential. We observe that all the experimental data increase monotonically with the density. We also
found that all microscopic EOSs are compatible with the experimental data. Although $E_{sym}$ obtained from original
TM1$^{*}$ and the TM1$^{*}$ with $\sigma$-cut potential follow the trend similar to other microscopic EOSs but both predict
high value of $E_{sym}$, around 5 MeV more compare to other EOSs after $\rho$ = $0.1$ $fm^{-3}$. There is no difference
in the value of $E_{sym}$ between original TM1$^{*}$ and the TM1$^{*}$ with $\sigma$-cut potential up to $0.3$ $fm^{-3}$.
We also found that $E_{sym}$ is softer at higher densities i.e. higher than $0.4$ $fm^{-3}$ in TM1$^{*}$
with $\sigma$-cut potential compared to original TM1$^{*}$. Such behavior of $E_{sym}$ occurs because TM1$^{*}$
with $\sigma$-cut potential ($f_{s}$ = 0.6) have a larger effective mass and in RMF models symmetry energy inversely
depend on effective mass\cite{Zhang:2018lpl}.    

In the present analysis, we consider the pure nucleonic as well as hyperon-rich neutron star matter. Meson-Hyperon couplings are fixed by reproducing the hyperon potentials at normal density as indicated by hypernuclei experiments. We fix the scalar $x_{\sigma H} = g_{\sigma H}/g_{\sigma N}$ and vector $x_{\omega H} = g_{\omega H}/g_{\sigma N}$ coupling constants by fixing the hyperon potential depths $(B/A)_{H}|_{\rho_{0}}$ as in ref. \cite{Sen:2018qvo} and is given by 

\begin{equation}
(B/A)_{H}|_{\rho_{0}} = x_{\omega H} U_{V}|_{\rho_{0}} + x_{\sigma H} U_{S}|_{\rho_{0}}
\label{hyp2}
\end{equation}
\noindent
where $U_{S} = g_{\sigma N}\sigma_{0}$ and $U_{V} = g_{\omega N}\omega_{0}$ respectively. We reproduce the hyperon potential 
depths $(B/A)_{H}$ at saturation density ${\rho_{0}}$ equal to -28MeV, +30MeV and -18MeV for $\Lambda$, $\Sigma$ and $\Xi$ respectively. We take $x_{\rho H} = x_{\omega H}$ in the present analysis. As suggested \cite{Glendenning:1991es}, the value of $x_{\sigma H}$ should be smaller than 0.72. In the present work, we take two values of $x_{\sigma H}$ = 0.6 and 0.7 and keep it the same for all hyperons. The corresponding vector coupling is taken by fixing $x_{\sigma H}$ and the aforementioned hyperon potentials at normal nuclear matter density. 

\begin{figure}
\includegraphics[width=8cm,height=8cm,angle=0]{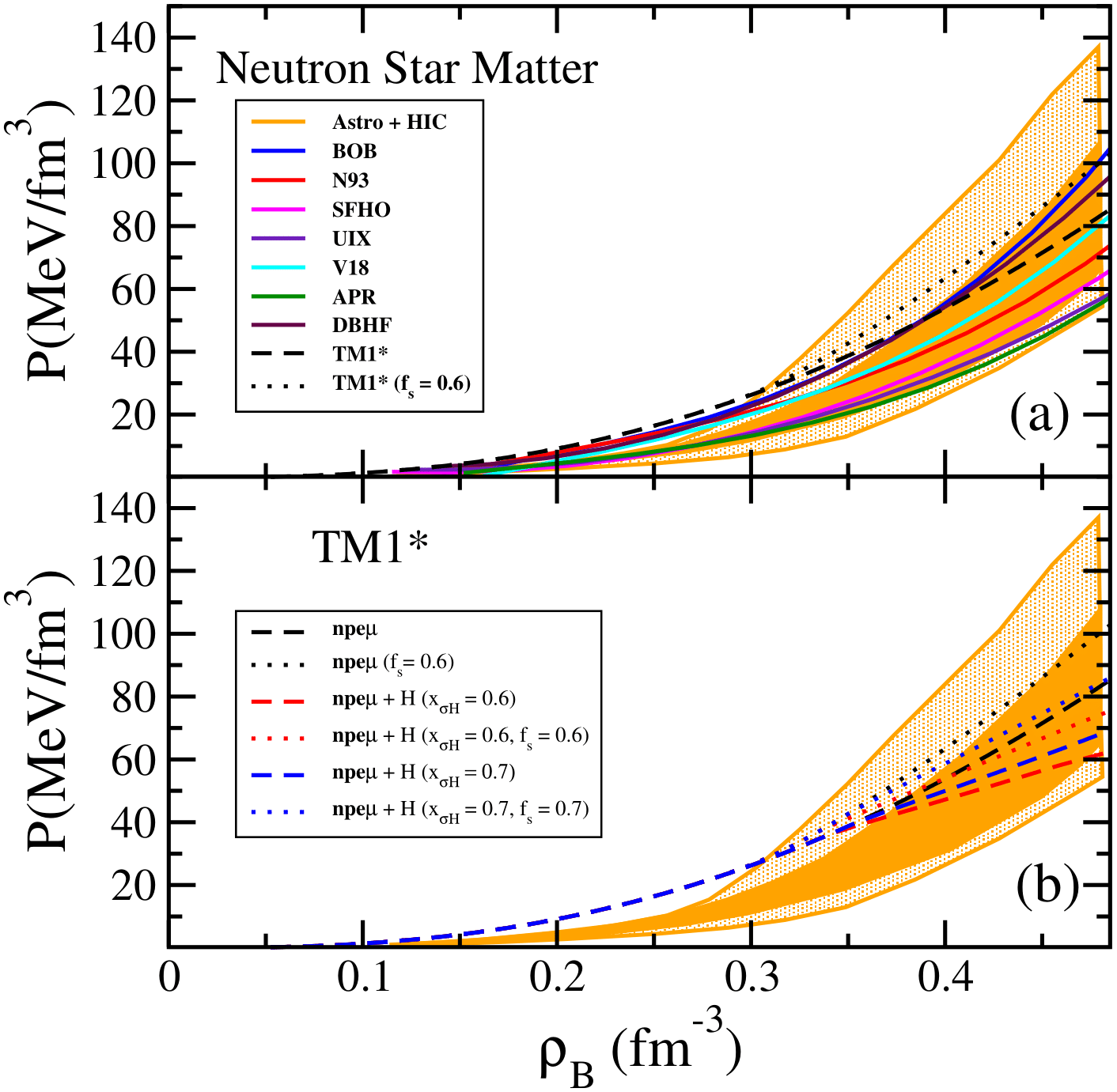} 
\caption{Pressure as a function of baryon number density $\rho_{B}$ for with or without $\sigma$-cut potential
in neutron star matter for (a) only nucleonic neutron star matter along with
other microscopic EOSs. and (b) neutron star matter including hyperons with original TM1$^{*}$ and
the TM1$^{*}$ with $\sigma$-cut potential
($f_{s}$ = 0.6) for scalar coupling constants $x_{\sigma H}$ = 0.6 and 0.7.
The combined data from multi-messenger neutron star observations and HIC (Astro+HIC) shown as an orange polygon is also shown. The shading corresponds to 95$\%$ and 68$\%$ credible intervals (lightest to darkest).}
\label{fig3}
\end{figure}
    
In Fig.3(a), we display the pressure as a function of baryon number density $\rho_{B}$ of neutron star matter.
As we are considering only nucleonic matter, the baryon number density $\rho_{B}$ is equal to $\rho$
like fig.1 and fig.2 in fig.3(a). The combined data from multi-messenger neutron star observations and
HIC (Astro+HIC) \cite{Huth:2021bsp}
shown as an orange polygon in both fig.3(a) and 3(b). The light and dark shade of the orange polygon corresponds to the 95$\%$ and 68$\%$ credible intervals, respectively. From Fig.3(a), one can conclude that the original TM1$^{*}$ and
the TM1$^{*}$ with $\sigma$-cut potential ($f_{s}$ = 0.6) along with other microscopic EOSs are consistent with
observational and experimental constraints for pure nucleonic neutron star matter. In Fig.3(b), we plot
the pressure as a function of baryon number density $\rho_{B}$ of neutron star matter with nucleonic and nucleonic plus hyperon rich matter for $x_{\sigma H}$ = 0.6 and 0.7. We found that all the EOSs are similar up to $\rho_{B} = 0.3
fm^{-3}$ and the effect of $\sigma$-cut potential and different $x_{\sigma H}$ starts appearing thereafter.
Our results for all EOSs in fig.3(b) are consistent with the dark shade area of combined data of observational and experimental constraints.

\begin{figure}
\includegraphics[width=8cm,height=8cm,angle=0]{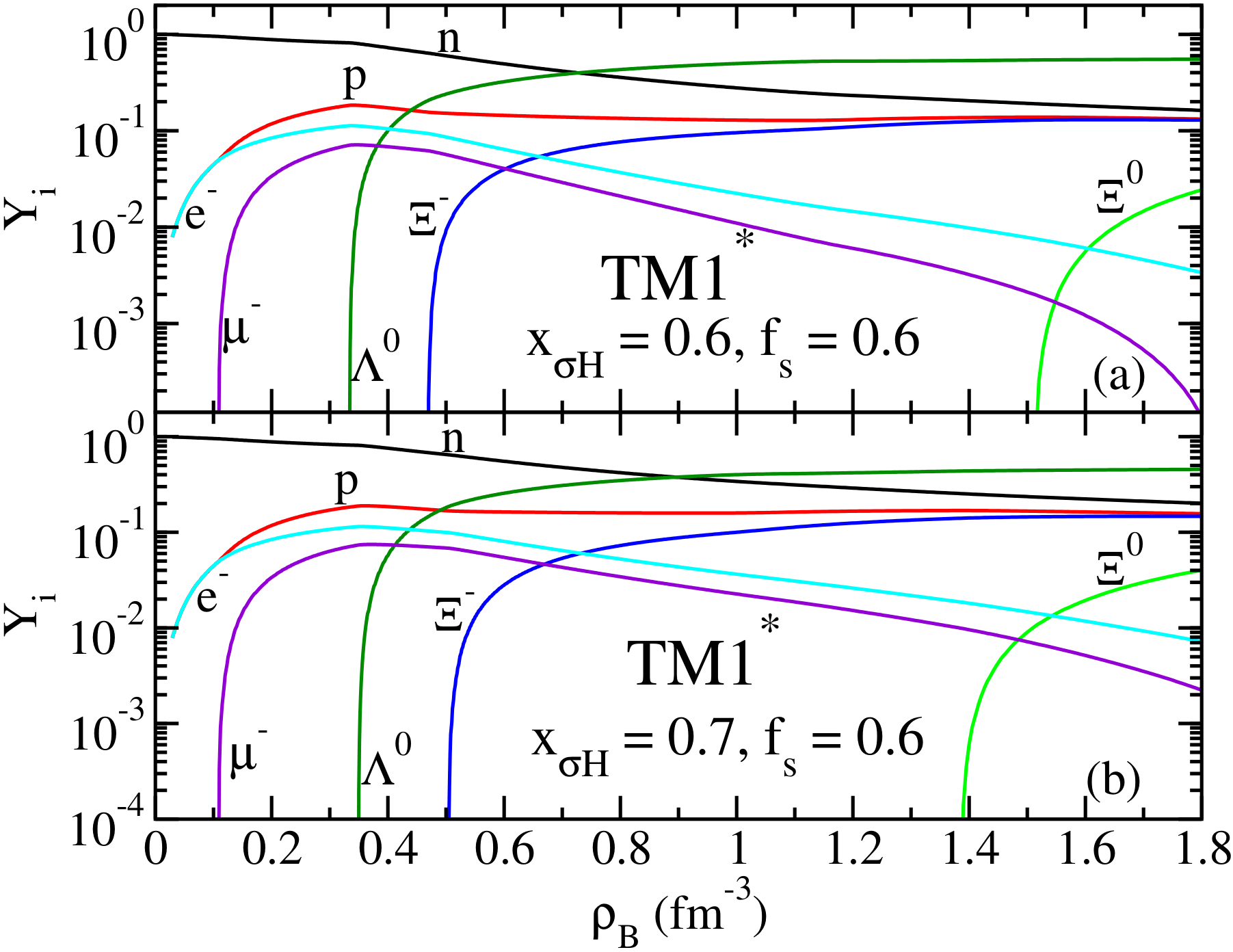} 
\caption{The particle fractions $Y_{i}$ as function of baryon number density $\rho_{B}$ 
in neutron star TM1$^{*}$ with $\sigma$-cut potential
($f_{s}$ = 0.6) for (a) $x_{\sigma H} = 0.6$ (b) $x_{\sigma H} = 0.7$.} 
\label{fig4}
\end{figure}

To analyse the effect of $\sigma$-cut potential on particle fractions $Y_{i} = \rho_{i}/\rho_{B}$,
we plot the particle fraction of TM1$^{*}$ with $\sigma$-cut potential ($f_{s}$ = 0.6)
for $x_{\sigma H} = 0.6, 0.7$ in fig.4(a) and (b) respectively. From fig.4(a) and 4(b), we find that
$\sigma$-cut potential has no effect on the neutron, proton, and lepton fractions (electron and muon) in the entire density range.
In the hyperonic sector, it is interesting to see that the $\sigma$-cut potential has a slight effect on the appearance of $\Lambda^{0}$ and $\Xi^{-}$. $\Lambda^{0}$ starts to nucleate at
 around $0.34 fm^{-3}$ for both the values of $x_{\sigma H}$. $\Xi^{-}$ appears at
$0.47 fm^{-3}$ for $x_{\sigma H} = 0.6$ and at $0.50 fm^{-3}$ for $x_{\sigma H} = 0.7$ respectively.
On the other hand, the main effect of $\sigma$-cut potential happens on $\Xi^{0}$ fractions
where $\Xi^{0}$ appears at $1.52 fm^{-3}$ for $x_{\sigma H} = 0.6$ as compared to $1.39 fm^{-3}$  
for $x_{\sigma H} = 0.7$.       

\begin{figure}
\includegraphics[width=8cm,height=8cm,angle=0]{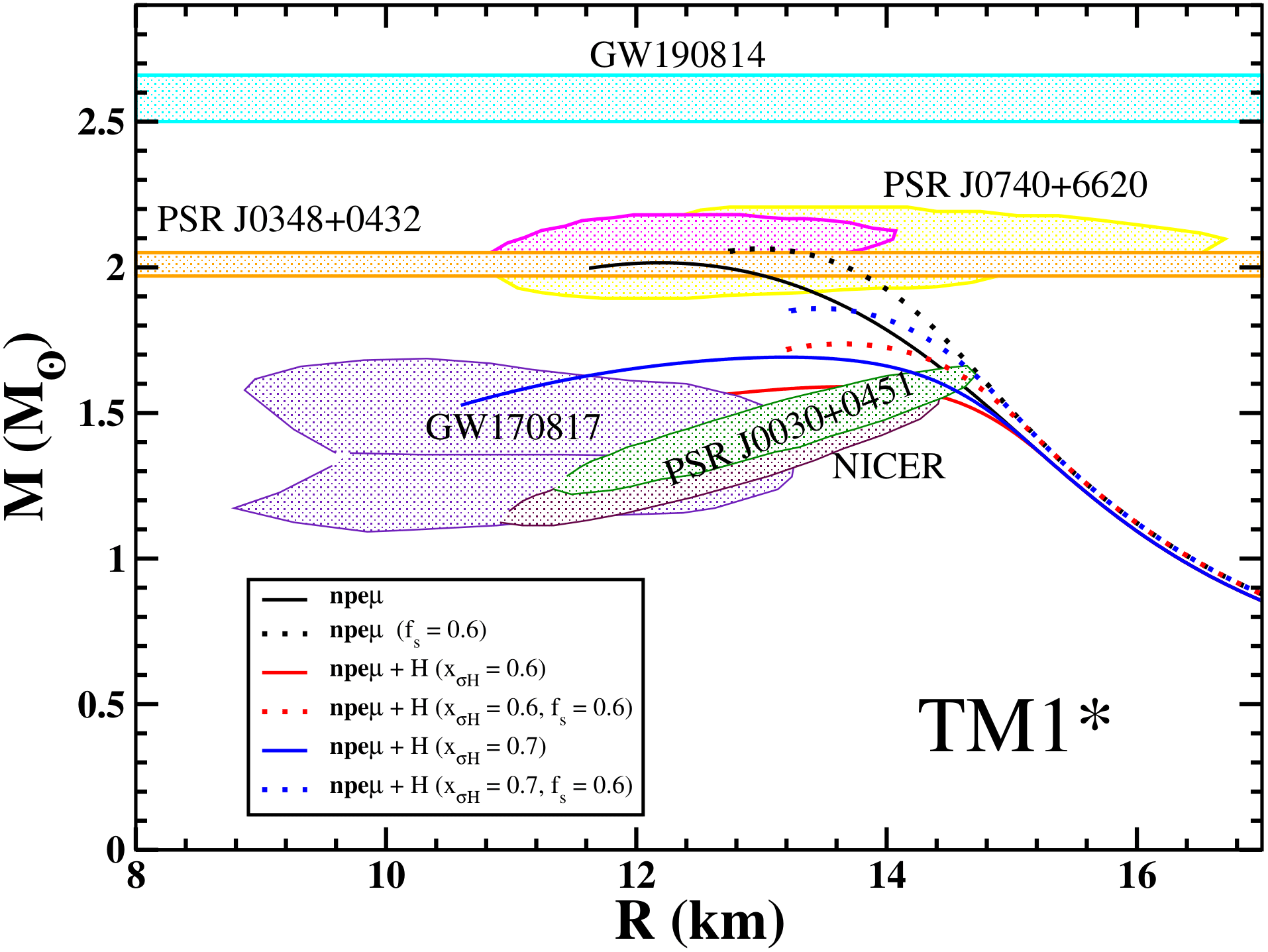} 
\caption{The mass radius relation of neutron star with original TM1$^{*}$ and TM1$^{*}$ with $\sigma$-cut potential for nucleonic and nucleonic plus hyperon rich matter with 
$x_{\sigma H}$ = 0.6 and 0.7 along with other available observational data.} 
\label{fig5}
\end{figure}

In Fig.5, we display the mass-radius (MR) relation for nucleonic and nucleonic plus hyperon-rich neutron star with original TM1$^{*}$ and TM1$^{*}$ with $\sigma$-cut potential 
($f_{s}$ = 0.6) along with observational data. The astrophysical observable constraints from 
GW190814 \cite{Abbott2020}, PSR J0740+6620 \cite{Riley:2021pdl,Miller:2021qha}, PSR J0348+0432 \cite{Antoniadis:2013pzd},
GW170817 \cite{Fattoyev:2017jql} and NICER experiment for PSR J0030+0451 \cite{Riley:2019yda,Miller:2019cac} 
are represented by shaded regions. As anticipated, the obtained mass 
for pure nucleonic matter ($n, p, e, \mu$) is larger compared to the mass obtained with hyperons 
for both the values of $x_{\sigma H}$. In hyperon-rich matter, mass is higher for 
$x_{\sigma H}$ = 0.7 compared to $x_{\sigma H}$ = 0.6 irrespective of whether we implemented
the $\sigma$-cut potential or not. The effect of $\sigma$-cut potential is almost similar
in all the considered TM1$^{*}$ EOSs. The TM1$^{*}$ with $\sigma$-cut potential gives 
the slightly higher mass of neutron star for nucleonic matter compared to the original TM1$^{*}$.

\begin{figure}
\includegraphics[width=8cm,height=8cm,angle=0]{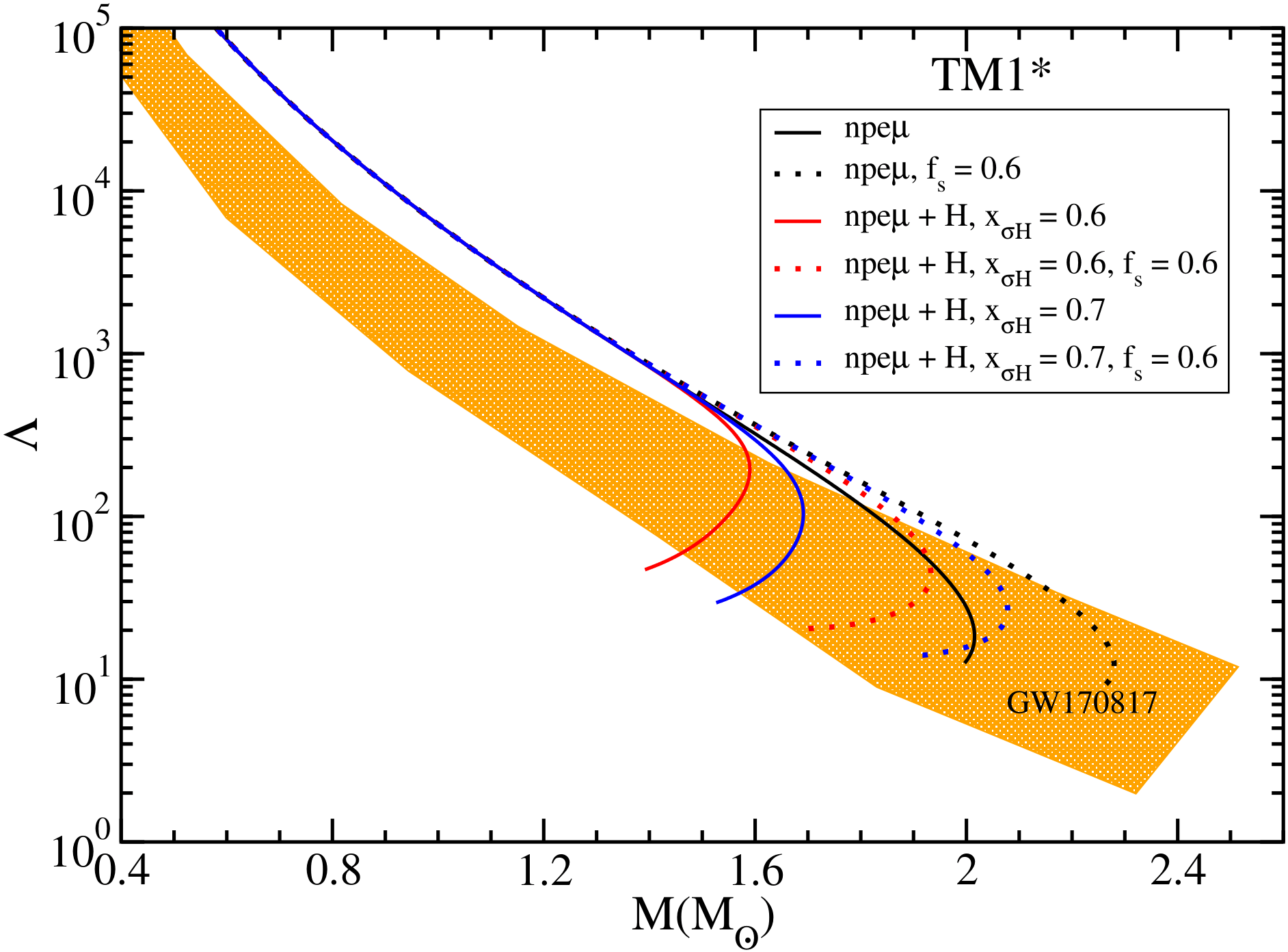} 
\caption{The tidal deformability of neutron star as function of maximum mass with 
original TM1$^{*}$ and 
TM1$^{*}$ with $\sigma$-cut potential for nucleonic and nucleonic plus 
hyperon rich matter with $x_{\sigma H}$ = 0.6 and 0.7 along with 
observational constraints from GW170817.} 
\label{fig6}
\end{figure}

In Fig.6, we display the calculated tidal deformability of neutron star with original TM1$^{*}$ and 
TM1$^{*}$ with $\sigma$-cut potential ($f_{s}$ = 0.6) for nucleonic and nucleonic plus 
hyperon rich matter with $x_{\sigma H}$ = 0.6 and 0.7. The observational constraints from
GW170817 is shown as a shaded region. Our results seem to be consistent with the observational constraints from GW170817. TM1$^{*}$ with 
$\sigma$-cut potential ($f_{s}$ = 0.6) for pure nucleonic matter has highest value of the
tidal deformability. Similarly, in hyperon-rich matter the value of tidal deformability 
is higher for $x_{\sigma H}$ = 0.7 compared to $x_{\sigma H}$ = 0.6. For comparison, all the calculated values of neutron star properties such as tidal deformability ($\Lambda_{1.4}$), radii (R$_{1.4}$ and R$_{2.07}$) (in km) and maximum mass (M$_{\rm max}$(in M$_\odot)$) with different TM1$^{*}$ models are listed in Table \ref{tab1}. It is concluded that with $\sigma$-cut potential ($f_{s}$ = 0.6) for pure nucleonic matter is well constrained and its maximum mass satisfy the current available maximum mass $2.08 \pm 0.07 M_\odot$ \cite{Miller:2021qha}. { Similar high mass neutron stars were obtained from three of other RMF models \cite{Dutra:2015hxa} investigated with nucleon only matter.}

\begin{table}[]
    \centering
    \caption{The values of neutron star properties, namely  the tidal deformability ($\Lambda_{1.4}$), radii (R$_{1.4}$ and R$_{2.07}$) (in km) and maximum mass (M$_{\rm max}$(in M$_\odot)$) are listed .\label{tab1}}
    \begin{ruledtabular} 
    \begin{tabular}{ccccc}
  TM1$^{*}$ models & M$_{max}$ & R$_{1.4}$ & R$_{2.07}$ & $\Lambda_{1.4}$\\[1.5ex]
  \hline
  npe$\mu$  & 2.02 & 15.24 & 12.18 & 845.30 \\[1.5ex]
  npe$\mu$, f$_s$=0.6  & 2.08 & 15.23 & 12.98 & 860.81 \\[1.5ex]
  npe$\mu$+H, x$_{\sigma H}$=0.6  & 1.59 & 15.13 & 13.67 & 841.74 \\[1.5ex]
  npe$\mu$+H, x$_{\sigma H}$=0.6, f$_s$=0.6   & 1.74 & 15.23 & 13.64 & 851.92 \\[1.5ex]
  npe$\mu$+H, x$_{\sigma H}$=0.7   & 1.69 & 15.12 & 13.20 & 831.66 \\[1.5ex]
  npe$\mu$+H, x$_{\sigma H}$=0.7, f$_s$=0.6   & 1.86 & 15.23 & 13.47 & 860.81 \\[1.5ex]
    \end{tabular}
    \end{ruledtabular}
\end{table}

\section{Conclusions and Summary} 

The $\sigma$-cut scheme developed \cite{Maslov:2015lma} and recently implemented to study neutron star properties of EOS with hyperons using TM1 parameter set \cite{Zhang:2018lpl} and kaon condensate using FSUGold parameter set \cite{Ma:2022fmu}. In the present analysis, we adopted and implemented the $\sigma$-cut scheme for the RMF model TM1$^{*}$ parameter, which gives the same results as TM1 for finite nuclei and nuclear matter properties \cite{DelEstal:2001yz}. We analyzed the effect of $\sigma$-cut potential on symmetric nuclear
matter, pure neutron matter, symmetry energy, and neutron star structure and composition with nucleonic and nucleonic plus hyperon-rich matter and compared with the existing experimental and observational data.
We found that $\sigma$-cut potential with TM1$^{*}$ parameter predicts a stiffer EOS compared to other microscopic EOSs for symmetric nuclear matter. A similar trend is also found in the case of pure neutron matter and symmetry
energy results. We found that there is almost no effect till
the two times nuclear saturation density $\rho_{0}$ and $\sigma$-cut potential make symmetry
energy softer at high density. We proceed to construct EOS with the original TM1$^{*}$
and TM1$^{*}$ with $\sigma$-cut potential ($f_{s}$ = 0.6) for pure nucleonic and nucleonic
and hyperon rich matter. The sensitivity of meson-hyperon coupling on the EOS is analyzed. For hyperon-rich matter, we choose scalar coupling constants
$x_{\sigma H}$ = 0.6 and 0.7, respectively, and fixed the respective hyperon potential depths in the matter as pronounced from the hyper-nuclear experiments. For $x_{\sigma H}$ = 0.6, we obtained smaller mass and tidal deformability as compared to $x_{\sigma H}$ = 0.7 of neutron
star. The dominant global properties of the neutron star, such as the mass, radius, and tidal deformability within the model, are in good agreement with experimental and observational
constraints.

\section{Acknowledgements} 
 N.K.P. would like to acknowledge the Department of Science and Technology, Ministry of Science and Technology, India, for the support of DST/INSPIRE Fellowship/2019/IF190058.
\newpage
%

\end{document}